\documentclass[useAMS,usenatbib,letterpaper]{mn2e}
\usepackage{times}
\usepackage{amssymb}
\usepackage{amsmath}
\usepackage{lscape,graphicx}

\title[Red sequence halo masses at $z > 2$]{Studying the emergence of the red sequence through galaxy clustering: host halo masses at $z > 2$}
\author[W. G. Hartley et al.]{W.~G. Hartley$^{1}$\thanks{E-mail:
    will.hartley@nottingham.ac.uk}, O. Almaini$^{1}$, A. Mortlock$^{1}$, C.~J. Conselice$^{1}$, R. Gr\"utzbauch$^{1,2}$, 
\newauthor C. Simpson$^{3}$, E.~J. Bradshaw$^{1}$, R.~W. Chuter$^{1}$, S. Foucaud$^{4}$, M. Cirasuolo$^{5}$, 
\newauthor J.~S. Dunlop$^{5}$, R.~J. McLure$^{5}$, H.~J. Pearce$^{5}$\\
  $^{1}$School of Physics and Astronomy, University of Nottingham, University Park, Nottingham NG7 2RD \\
  $^{2}$Center for Astronomy and Astrophysics, University of Lisbon, Portugal \\
  $^{3}$Astrophysics Research Institute, Liverpool John Moores University, Twelve Quays House, Egerton Wharf, Birkenhead CH41 1LD  \\
  $^{4}$Department of Earth Sciences, National Taiwan Normal University, No. 88, Section 4, Tingzhou Road, Wenshan District, Taipei 11677, Taiwan \\
  $^{5}$Institute for Astronomy, University of Edinburgh, Royal Observatory, Edinburgh EH9 3HJ \\
}

\begin{document}

\date{}

\pagerange{\pageref{firstpage}--\pageref{lastpage}} \pubyear{2012}

\maketitle

\label{firstpage}

\begin{abstract}

\noindent We use the UKIDSS Ultra-Deep Survey, the deepest degree-scale 
near-infrared survey to date, to investigate the clustering of 
star-forming and passive galaxies to $z \sim 3.5$. Our new measurements include 
the first determination of the clustering for passive galaxies at $z>2$, 
which we achieve using a cross-correlation technique. We find that passive 
galaxies are the most strongly clustered, typically hosted by massive dark matter 
halos with M$_{\rm halo} > 5\times 10^{12}~$M$_{\odot}$ irrespective of redshift or stellar 
mass. Our findings are consistent with models in which a critical halo 
mass determines the transition from star-forming to passive galaxies. 
Star-forming galaxies show no strong correlation between stellar mass and 
halo mass, but passive galaxies show evidence for an anti-correlation; low-mass passive 
galaxies appear, on average, to be located in the most massive halos. 
These results can be understood if the termination of star formation is 
most efficient for galaxies of low stellar mass in very dense 
environments.

\end{abstract}


\section{Introduction}
\label{intro}

One of the earliest and most fundamental observational results in extragalactic astrophysics is the bimodal form of the galaxy population. Ever since the time of \cite{Holmberg58}, it has been clear that galaxies can be separated into two broad classes based on little more than their rest-frame optical colours. A lack of recent star-formation is responsible for the red colours of many galaxies, as the massive stars that emit light at shorter wavelengths are short lived. We are yet to fully determine, however, why star formation should cease in some galaxies yet continue in others.

The vast majority of relatively non-star-forming (hereafter, passive) galaxies are also observed to have spheroidal morphologies, while local star-forming galaxies predominantly show disks \citep[e.g.][]{Kennicutt98}. A natural approach is therefore to ask whether the same process responsible for morphological transformation is also responsible for the termination of star formation. \cite{Toomre77} suggested that the merger of two similar-mass disky systems could result in a spheroid. A major merger, or some other bulge-forming event, remains among the most favoured explanations to date for the transition to a passive state \citep[e.g.][]{Bell11}. In this scenario a gas-rich merger or collapse of an unstable disk results in very vigorous star formation and possibly a quasar phase. Either of these could have sufficient energy to expel the remaining gas from the galaxy, or heat it sufficiently to prevent further star formation \citep{Silk98,Fabian99}. Without a reservoir of cold gas the galaxy can no longer form stars and it is observed as a passive object. Episodic AGN feedback may then be required to prevent the gas from re-cooling onto massive galaxies \citep[e.g.][]{Benson03,Best06}.

Theoretical considerations suggest an additional requirement for galaxies to end their star formation. Expanding on the early work of \cite{White78}, recent galaxy-formation models predict a declining rate of star formation in galaxies that depends on the virial temperature of the dark matter halo in which the galaxy is located \citep{Croton06,Dekel09,Feldmann10,Cen11}. This so called `hot halo' model predicts that as a dark matter halo grows in mass it becomes able to sustain a reservoir of gas with long ($\sim$ few Gyrs) cooling times. New gas infall is shock heated as it encounters this gas halo and is therefore not immediately available for star formation. Henceforth, only a modest amount of energy input, possibly from radio AGN feedback, is required to maintain a high gas temperature. In this way, galaxies within high mass halos become starved of star-forming gas with the result that ongoing star formation rates become negligible. This theory explains the observed trends of downsizing: the most massive galaxies are the first with halos of sufficient mass to suppress star formation \citep{Cowie96,Kodama04}. A clear consequence of the hot halo model is that there should be a characteristic halo-mass scale, above which the relative fraction of passive objects (at fixed stellar mass) increases sharply. Current models suggest that this halo-mass scale is a few$~\times 10^{12}$M$_{\odot}$ and becomes maximally efficient at M$_{{\rm halo}}\sim 10^{13}~$M$_{\odot}$ \citep{Croton06,Cen11}. This mass scale should not depend on the stellar mass of the galaxy sample, nor on the redshift of the objects.

Estimating the masses of dark-matter halos that host passive galaxies is important to test these theoretical predictions. However, obtaining such measurements for individual systems is challenging, even in cases where excellent spectroscopic and high-resolution data are available. Direct halo-mass estimates (e.g. via gravitational lensing or velocity dispersions) are typically only available at low to intermediate redshifts, and only reliable for higher-mass halos \citep*{Mandelbaum09}. Systematically studying the host halo masses of large populations at high redshift is currently beyond these techniques. Fortunately our knowledge of the background cosmology in which galaxies form and evolve is now sufficiently advanced \citep{Komatsu11} such that we can describe the growth of the dark matter framework with reasonable confidence \cite[e.g.][]{Sheth99,Mo02,Springel05}. This confidence has in turn allowed us to link galaxy samples to a typical dark matter halo mass via the use of their clustering statistics on linear scales \citep{Peebles80}.

In the low-redshift Universe, surveys such as the SDSS \citep{York00} and 2DFGRS \citep{Colless01} have enabled studies of vast galaxy populations that had not been possible previously. Clustering studies based upon these data routinely find red or early type galaxy samples (those that are dominated by passive objects) cluster more strongly than an equivalent (by stellar mass or luminosity) star-forming sample \citep[e.g.][]{Norberg02, Zehavi05, Ross09, Zehavi11}. Moreover, contrary to star-forming samples, the clustering strength of passive galaxies does not appear to increase with luminosity \citep{Norberg02, Zehavi05}. Often, the stronger clustering of passive galaxies is attributed to a higher satellite fraction in the red or passive sample \citep{Zehavi05, Ross09}. We return to this point in section \ref{disc}. However, in order to fully explore the mechanisms that cause a galaxy to cease its star formation, we must study the population as the transition occurs, i.e. at higher redshift.

At intermediate redshift ($z<1$) a number of studies have sought to determine whether the difference between passive and star-forming clustering strengths continues. Optically-selected surveys, such as DEEP2 \citep{Davis03}, VVDS \citep{LeFevre05} and CFHTLS, can be used up to $z\sim1$, but become heavily biased against red objects at higher redshift. Studies using these data find very similar results to those in the low-redshift Universe \citep{Meneux06,Coil08,McCracken08,Coupon11}: red or passive samples are, on average, found in more massive dark matter halos than equivalent star-forming objects. In fact this clustering strength depends not only on whether the sample is star-forming or passive, but continuously as a function of colour \citep{Coil08}.

In order to use the large-scale structure to estimate halo masses at $z>1$ we require selection at near-infrared wavelengths. A large contiguous field is clearly necessary to measure the bias at large scales (r$~\sim~$few Mpc) and find rare objects at high redshift, but equally important is excellent depth covering a wide range in wavelength. There have been a number of studies using a variety of data and colour-selection techniques to investigate the clustering of passive and star-forming galaxies at $z>1$ \citep{Roche02,Kong06,Hartley08,Kim11}. However, it is only in recent years that survey data of the depth and breadth in wavelength required to construct purely mass or luminosity-limited samples have become available. Arguably the leading near-infrared survey for such work is the UKIDSS Ultra Deep Survey (UDS, see section \ref{data}) and these data have been used to produce two studies of particular relevance to this work. 

\cite{Williams09} used the UDS data release 1 (DR1) to identify passive and star-forming galaxy samples based on the rest-frame colour selection criteria first introduced by \cite{Wuyts07}. Comparing the clustering strengths of the two samples over the redshift range $1<z<2$, they found that passive galaxies are hosted by more massive halos. Breaking each sample into two luminosity-dependent sub-samples, they further found that there was no luminosity dependence on halo mass in the passive sample. This was interpreted as evidence for a common, or limiting, halo mass for passive galaxies at this redshift. 

In \cite{Hartley10} (hereafter, H10) we performed a more extensive analysis using the third data release (DR3), over the full redshift range accessible from the data. We split star-forming and passive galaxy samples into subsamples by redshift and K-band luminosity and computed their clustering statistics. The results confirmed and expanded upon those of \cite{Williams09}, finding a roughly constant halo mass for all passive subsamples, but clear evolution and generally lower halo masses for star-forming sub-samples. Due to the small sample of passive galaxies at $z>2$ however, we were unable to compute a typical halo mass for passive galaxies at this redshift. However, the trends shown by the two populations suggested that passive and star-forming galaxies would be found in equal mass halos by $z\sim2.5$. 

The latest UDS data are approximately 1 magnitude deeper in all bands compared to those used in H10, enabling us to investigate the host halo masses of a stellar-mass-selected population of passive galaxies at $z>2$ for the first time. Furthermore, we can probe the environments of lower-mass passive objects (M$_{*} < 10^{10}~$M$_{\odot}$) to $z\sim1.5$. Low-mass passive galaxies at these redshifts are extremely faint but are a potentially important population for discriminating between galaxy formation models. Before the measurements presented in this work there were no estimates of the typical host halo mass for such low-mass galaxies to our knowledge. With these two vital pieces of information together with a consistent set of measurements in three stellar-mass bins across $0.3<z<3.36$ we re-visit whether there is a possible host halo-mass limit necessary to suppress star-formation in galaxies and produce passive objects.

The structure of this paper is as follows: in section~\ref{data} we discuss the data sets employed in this work and our sample construction; section~\ref{method} details our methodology and principal results, which are then discussed in section~\ref{disc}. We conclude in section~\ref{conc}. Throughout this work, where appropriate, we adopt a $\Lambda$CDM cosmology with $\Omega_M = 0.3$, $\Omega_{\Lambda} = 0.7$, h$ = {\rm H}_0/100 ~{\rm kms}^{-1}{\rm Mpc}^{-1} = 0.7 $ and $\sigma_8 = 0.9$. All magnitudes are given in the AB system \citep{Oke83} unless otherwise stated. Finally, the definition of `halo mass' varies in the literature; in this work we are referring to the parent halo in which a galaxy resides, as opposed to the mass of its immediate sub-halo.

\section[]{Data sets, derived quantities and sample definitions}
\label{data}

\begin{figure}
\begin{center}
\includegraphics[angle=0, width=250pt]{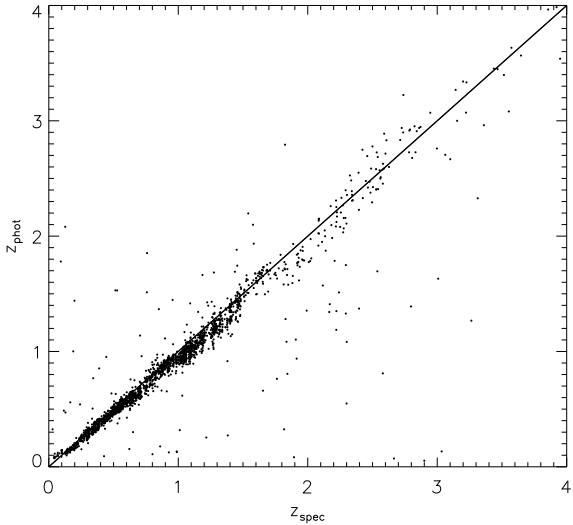}
\caption{Probability-weighted photometric redshifts versus spectroscopic redshifts for the combined UDSz and archival redshift samples ($2147$ galaxies). The dispersion ($\Delta z/(1+z)$) and outlier fraction are $0.031$ and $3.3\%$ respectively.}
\label{zspec}
\end{center}
\end{figure}

Our sample is drawn from the K-band image of the UKIRT Infrared Deep Sky Survey (UKIDSS, \citealt{Lawrence07}), Ultra-Deep Survey\footnote[1]{http://www.nottingham.ac.uk/astronomy/UDS/} (UDS) data release 8 and matched multi-wavelength photometry. The UDS is the deepest of the five UKIDSS sub-surveys, consisting of four Wide-Field Camera (WFCAM, \citealt{Casali07}) pointings, covering $0.77$ square degrees in J, H and K-bands. Observations of the ongoing UKIDSS survey began in the spring of 2005 and the UDS DR8 comprises all UDS data taken up to and including January 2010. Total on-source exposure times in the three near-infrared bands are 186.5, 100 and 207.5 hours for J, H and K respectively, reaching depths of ${\rm J}=24.9$, ${\rm H}=24.2$ and ${\rm K}=24.6$ (AB magnitudes, estimated from the RMS between $2\arcsec$ apertures). The combined exposure time of these data is more than double that of the previous release, resulting in the deepest near-infrared images over a single degree-scale field to date. In the case of our selection band (K-band), the data are more than $0.5$ magnitudes deeper than in any comparable field. Details of the stacking procedure, mosaicing, catalogue extraction and depth estimation will be presented in Almaini et al. (in prep.). For this work we perform a catalogue extraction from the K-band image using SExtractor. We produced a merged source list based on two catalogues: the first was designed to efficiently de-blend sources and find compact objects; while the second was optimised for extended but low-surface brightness galaxies. These are often referred to as `hot' and `cold' catalogues in the literature \citep[e.g][]{Rix04}.

In addition to the three UKIDSS bands, the field is covered by comparable data from CFHT Megacam u-band, optical Subaru Suprime-cam data and Spitzer IRAC channels 1 and 2. The u$^{\prime}$-band data reach ${\rm u^{\prime}} = 26.75$ (AB, $2\arcsec$ RMS) and are detailed in Foucaud et al. (in prep.). Deep optical data in the B, V, R, i$^{\prime}$ and z$^{\prime}$-bands are taken from the Subaru-XMM Deep Survey (SXDS), achieving depths of ${\rm B} = 27.6, ~{\rm V} = 27.2, ~{\rm R} = 27.0, ~{\rm i}^{\prime} = 27.0$ and ${\rm z}^{\prime} = 26.0$ \citep[$5\sigma$, $2\arcsec$]{Furusawa08}. Data at longer wavelengths are crucial for accurate stellar-mass measurements of galaxies at high redshift. The UDS {\it Spitzer} Legacy Program (SpUDS, PI:Dunlop) provides deep data in channels 1 and 2 of IRAC, as well as MIPS $24\mu m$ data, all of which are used during our analysis. SpUDS data reach $5\sigma$ depths of $24.2$ and $24.0$ (AB) at $3.6\mu$m and $4.5\mu$m respectively, while the public $24\mu$m catalogue used here is limited to $300~\mu$Jy ($15\sigma$). The co-incident area of these data sets after masking of bad regions and bright stars is $0.62$ square degrees. Finally, we make use of existing X-ray \citep{Ueda08} and radio \citep{Simpson06} data to remove obvious AGN.

Photometry was extracted in $3\arcsec$ apertures placed on each astrometrically aligned image at the position of the K-band sources \citep[see][for further details]{Simpson12}. Due to correlated noise in the images that is not represented in the weight maps, magnitude uncertainties are underestimated by SExtractor. We account for this correlated noise by scaling the weight maps such that the uncertainty in source-free regions matches the RMS measured from apertures placed on the science image. 

Three of the bands (the CFHT u$^{\prime}$-band and the two IRAC channels) required PSF corrections to their photometry in order to obtain correct colours. This correction was performed as follows. For the IRAC channels, the K-band image was used to estimate the flux lost from the aperture due to the broad IRAC PSF. The K-band image was smoothed to $2.0\arcsec$ FWHM (the FWHM of the IRAC images) and the ratio of counts within $3\arcsec$ apertures in the original un-smoothed image and the smoothed image was computed for each source. The IRAC fluxes were then multiplied by this ratio. In addition to correcting for the difference in PSF, this method automatically accounts for contamination in the $3\arcsec$ IRAC fluxes by neighbouring sources. The caveat is, however, that the source and contaminating flux are assumed to have the same K$-$IRAC colour. We discuss this caveat further in section~\ref{disc}. The IRAC PSF correction is typically of order $26-30\%$ of the original flux. The u$^{\prime}$-band fluxes were corrected in a similar way using the B-band image, however the B and u$^{\prime}$-band PSFs are similar and so the typical correction is less than $1\%$ for this band.

\subsection{Photometric redshifts}
\label{photz}

\begin{figure}
\begin{center}
\includegraphics[angle=0, width=240pt]{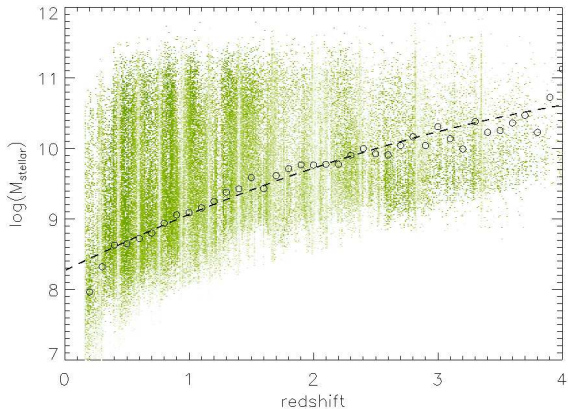}
\caption{Log(stellar mass) versus photometric redshift for our galaxy sample (green) where the colour corresponds to the galaxy weight (paler colours indicate lower weights). Black circles indicate the $95\%$ mass completeness limit at that redshift (see text), while the dashed line is a second order polynomial fit to these points.}
\label{Mvsz}
\end{center}
\end{figure}

At the depths probed in our study it is essential to use photometric redshifts ($z_{{\rm phot}}$). In addition to extremely deep photometry, the UDS field was also the target of a unique spectroscopic survey: the UDSz. The UDSz is based on an ESO Large Programme targeting a large sample of galaxies ($\sim3500$) at $z_{{\rm phot}}>1$ with ${\rm K}_{{\rm AB}}<23.0$, plus a low-redshift control sample. The survey comprised eight pointings of VIMOS in LR-Blue and LR-Red and 20 FORS2 masks with the GRS\_300I grating. This survey has produced $\sim 1500$ secure redshifts to date which are used along with archival redshifts (of $\sim4000$ objects), after excluding objects with clear active galactic nuclei (AGN), to train our photometric redshifts. The UDSz data will be described fully in a forthcoming paper (Almaini et al., in prep.) and a data release is expected to follow in late 2013. Details of the archival redshifts can be found in \cite{Simpson12} and references therein.

Photometric redshift probability distributions were computed for each source using the EAZY package \citep{Brammer08} with the included apparent K-band magnitude prior, after correcting the observed fluxes for Galactic extinction (following the maps of \citealt{Schlegel98}). The default set of six templates does not sufficiently represent all of our galaxies; in particular the u$^{\prime}$-band flux of blue objects at high redshift is significantly overestimated. We therefore constructed a seventh template, applying a small amount of SMC-like extinction \citep{Prevot84} to EAZY's bluest template. The differences between the observed and template fluxes were then used to iteratively correct the photometric zero points with the following adjustments relative to the i$^{\prime}$-band required: ${\rm u^{\prime}}=-0.15$, $3.6\mu{\rm m}=+0.04$ and $4.5\mu{\rm m}=+0.05$ (where a negative number means the observed magnitudes must be made brighter). These systematic corrections differ slightly from those in \cite{Caputi11}. However, the addition of the UDSz spectra produces an overall spectroscopic sample which is more representative of the K-selected photometric sample used here than previously available samples in the field. The final photometric redshifts of the subset with spectra (UDSz plus archival) are compared with their spectroscopic redshifts in Figure~\ref{zspec}. We have excluded bright X-ray sources and those with clear AGN signatures in their spectra, but a number of AGN are likely to remain in these samples. The dispersion of $z_{{\rm phot}}-z_{{\rm spec}}$ for these redshifts after excluding outliers ($\Delta z/(1+z)>0.15$, $<4\%$ of objects) is $\Delta z/(1+z) \sim 0.031$. Most previous photometric redshift sets within the field have been tested against the archival spectra only, which are heavily weighted towards $z=0$. As we describe in the next section, the full probability distribution produced by EAZY, rather than a single probability-weighted redshift, is used throughout our analysis.

\subsection{Parent sample construction}
\label{samplecons}

We now describe the preparation of the parent galaxy sample from which our passive and star-forming samples are drawn. Firstly, a number of cuts were made to ensure a clean but complete dataset. Objects were identified as stars and removed if they satisfied one of the following conditions: 
\begin{itemize}
\item For objects with $K < 20$, bright stellar objects are easily identified from a locus in 
light profiles defined by $K_{1.5\arcsec}-K_{0.7\arcsec}$. We make additional constraints that stars lie on the stellar locus in $J-K$ vs $K$ and in the stellar locus of a K-excess diagram ($V-J$ vs $J-K$). Saturated objects on the K-band image (typically $K<14$) are all assumed to be stars.
\item Objects with $K > 20$ are required to simultaneously lie in the stellar locus of a K-excess 
diagram, a $J-K$ vs $K$ diagram and a $BzK$ diagram.
\end{itemize}
These criteria are deliberately conservative as we do not wish to remove any possible compact galaxies. There are $3806$ objects identified as stars in this manner. A second source of possible contamination in the galaxy sample are spurious sources produced by amplifier cross-talk \citep{Dye06}. For cross-talk produced by bright stars, visual identification is trivial and these are masked along with the stars that produce them. However, even faint objects produce cross-talk in deep stacks such as the UDS and these require much greater care to remove. The positions of cross-talk artifacts can be predicted, as they occur at regular intervals ($128$ native WFCAM pixels) both horizontally and vertically from the source that produced them. Many cross-talk exhibit positive and negative pairs and these cases can be confirmed by cross-matching the predicted positions with a $5\sigma$ source list from an inverted image. Furthermore, as the optical Subaru images do not suffer from cross-talk contamination, the optical-near infrared colours can be used to select a list of likely artifacts. All candidate cross-talk sources from these two methods were then inspected in the K-band image to ensure that we were not removing any real objects. We find a total of $3419$ cross-talk artifacts\footnote[3]{NB. A less conservative stellar catalogue was used to predict the positions of cross-talk.}.

From the resulting galaxy sample a total K-band magnitude (SExtractor's MAG\_AUTO) limit was imposed at ${\rm K}=24.3$ (AB). From simulation of sources we have determined that our catalogue is $>99\%$ complete at this limit (Almaini et al. in prep.) and it ensures that slight variations in the depth across the UDS image do not affect our measurements. Following this, we also impose a photometric redshift quality cut. Any galaxy in which the minimum $\chi^2$ during $z_{{\rm phot}}$ fitting was greater than $11.35$ was excluded. Those with high $\chi^2$ cannot be assigned a reliable probabilty of being at any redshift and will therefore contaminate our samples. Many of these objects contain AGN or are blended source detections, both of which alter the galaxies' photometry and prevent us from assigning them to an appropriate star-forming or passive galaxy sample. In addition we removed likely AGN by excluding objects associated with either a point-source X-ray detection or radio source brighter than $100\mu m$. Combined, these cuts remove $15\%$ of our sample. We repeated our measurements with these criteria relaxed, excluding only those with $\chi^2>100$ ($<1\%$ of the sample) but found qualitatively identical results. We present only those mesurements from our cleaner sample in this paper.

Our measurements are carried out in redshift slices, $500$Mpc (physical) in depth. This depth corresponds to at least three times the photometric redshift dispersion quoted in section~\ref{photz}. Rather than assign each galaxy to a particular redshift, we make use of the full probability distribution and allow a galaxy to be represented multiple times. For each redshift slice, $z1<z<z2$, the integrated probability of a galaxy being within that redshift range is computed: $p_{{\rm slice}} = \int^{z2}_{z1} p(z) dz$. Provided that $p_{{\rm slice}} > 0.02$, an entry is made in the catalogue for that galaxy and redshift. This probability limit is imposed to reduce the computational cost of our measurements. For all statistical measures presented in this work, the integrated probability is used as a weight. Stellar masses and rest-frame colours (see below) are computed separately for each entry, at the minimum $\chi^2$ redshift within the interval. For example, a galaxy may have a broad peak in redshift probability, contained within two redshift slices, and a separate minor solution at some other redshift. This galaxy would be represented three times in our measurements, with three different values for stellar mass and three sets of rest-frame colours. A similar strategy to ours was employed by \cite{Wake11}. For simplicity we shall henceforth refer to each unique object and redshift as a ``galaxy''.

\subsection{Stellar masses, M$_*$}
\label{masses}

The stellar masses and rest-frame luminosities used in this work are measured using a multicolour stellar population fitting technique. We fit the $u^{\prime}BVRi^{\prime}z^{\prime}JHK$ bands and IRAC Channels 1 and 2 to a large grid of synthetic spectral energy distributions (SEDs) constructed from the stellar population models of \cite{Bruzual03}, assuming a Chabrier initial mass function. The star-formation history is characterised by an exponentially declining model with various ages, metallicities and dust extinctions. These models are parametrised by an age since the onset of SF, and by an e-folding time such that
\begin{equation}
SFR(t) \sim SFR_{0}\times e ^{-\frac{t}{\tau}}.
\label{eq:SFRexp}
\end{equation}
where the values of $\tau$ can range between 0.01 and 13.7 Gyr, while the age of the onset of star formation ranges from 0.001 to 13.7 Gyr. We exclude templates that are older then the age of the Universe at the redshift of the galaxy being fit. The metallicity fraction is allowed to range from 0.0001 to 0.1, and the dust content is parametrised, following \cite{Charlot00}, by $\tau_{v}$, the effective V-band optical depth. We use values up to $\tau_{v}=2.5$ with a constant inter-stellar medium fraction of $0.3$.

To fit the SEDs they are first scaled in the observed frame to the K-band magnitude of the galaxy. We then fit each scaled model template in the grid of SEDs to the measured photometry of the galaxy. We compute $\chi^2$ values for each template and select the best-fitting template, obtaining a corresponding stellar mass and rest-frame luminosities. 

From the catalogue of stellar masses a $95\%$ mass completeness limit as a function of redshift was found following \cite{Pozzetti10}. At each redshift, the galaxies within $\Delta z=0.05$ of the target redshift were identified and sorted by their total (observed) K-band magnitude and the faintest $20\%$ selected. Each of these objects then had its stellar mass scaled to the value that it would have if its magnitude were equal to our imposed limit (${\rm K}=24.3$). The $95\%$ mass completeness limit was taken as the $95^{{\rm th}}$ percentile of the resulting scaled mass distribution. These redshift-dependent mass limits are shown in Figure \ref{Mvsz} by small black circles. Also shown are a polynomial fit to the limits (dashed line, ${\rm M}_{{\rm lim}}=8.27+0.87~z-0.07~z^2$) and our galaxy sample (green points). Darker points indicate greater weights. Galaxies which fall below ${\rm M}_{{\rm lim}}$ are not used in the subsequent analysis.

\subsection{Sample selection}
\label{sample}

The division of our sample into passive and star-forming subsets could be performed in a number of ways. Here we are interested in those galaxies which have completed their major episode of star-formation and therefore show evolved stellar populations and very low residual star-formation. Our selection is based initially on the U, V and J-Bessel band rest-frame luminosities. These were used by \cite{Williams09} to separate evolved stellar populations from those with a greater fraction of recent star-formation (see also \citealt{Wuyts07}). Due to the small sample of high redshift ($z>2$) passive galaxies, \cite{Williams09} were only able to find a clear passive sequence to $z\sim2$. Calibration of passive selection criteria at higher redshift would require spectra which we do not have. However, our data are 1 magnitude deeper than those in \cite{Williams09} and the passive sequence does appear to continue to higher redshift, as predicted by synthetic evolution models. We therefore take a simple extension of the \cite{Williams09} selection criteria to higher redshift. For completeness we list the colour criteria derived by \cite{Williams09} to select passive objects that we use in this work:
\begin{align*}
{\rm U-V} > 0.88\times{\rm V-J} + 0.69 && (z<0.5) \\
{\rm U-V} > 0.88\times{\rm V-J} + 0.59 && (0.5<z<1.0) \\
{\rm U-V} > 0.88\times{\rm V-J} + 0.49 && (z>1.0)
\end{align*}
with ${\rm u-V} > 1.3$ and ${\rm V-J} < 1.6$ in all cases. The remaining objects are assigned to our star-forming sample. Although these criteria efficiently select galaxies with predominantly old stellar populations, there is a possibility that the sample could still be contaminated by dusty star-forming objects, edge-on discs or AGN. Furthermore, the deep broadband photometry available across the electromagnetic spectrum provides much more information than is used in a simple UVJ selection. We therefore use the template fits obtained during stellar-mass determination to estimate the specific star-formation rate (SSFR) of each catalogue member and identify clear cases of dusty star-forming contaminants in the passive sample. Objects with passive UVJ colours but ${\rm SSFR} > 10^{-8} ~{\rm yr}^{-1}$ were re-assigned to the star-forming sample. Furthermore, the $24\mu {\rm m}$ data can also be used to identify red objects that harbour dust-enshrouded star-formation. The SFR can be estimated from the mid-infrared flux by
\begin{align*}
{\rm SFR} ({\rm M}_{{\odot}} {\rm yr}^{-1}) = 7.8\times10^{-10} ~{\rm L~(24\mu m, L}_{{\odot}})
\end{align*}
\citep{Rujopakarn11}. We took a simplistic approach to $24\mu m$ source identification, assigning each $24\mu$m source to its closest K-band counterpart within $6\arcsec$ (though the vast majority were matched within $2\arcsec$). For matching objects, the $24\mu$m flux was converted to a luminosity (without K-correction), based solely on its redshift, and then to a SSFR as above. Any passive galaxy with SSFR$_{24\mu {\rm m}} > 7.43\times 10^{-11} ~{\rm yr}^{-1}$ (i.e. a stellar-mass doubling time less than the $z=0$ Hubble time for our cosmology) was also re-assigned to the star-forming sample. In total $\sim3\%$ of passive objects selected via the UVJ criteria were moved to the star-forming sample, $90\%$ of these were due to a $24\mu$m detection. 

Given the small fraction of galaxies reassigned to the star-forming sample, it is unsurprising that relaxing these criteria has minimal impact on our results. In addition we have tested samples in which a stricter cut in the SSFR implied by the template fit is made. The results are, again, very similar to those presented here, but in making a stricter cut the risk of reassigning genuine passive galaxies to the star-forming sample is increased. We therefore present the results for only the samples that we believe provide our cleanest separation between passive and star-forming galaxies. We note that the construction of our passive and star-forming samples differ from our earlier work (H10). In H10 we contrasted the extremes of the star-forming and passive samples. In this work our aim is to achieve much more complete samples and we therefore place no requirement on star-forming galaxies to be vigorously star-forming.

\begin{figure}
\includegraphics[angle=0, width=240pt]{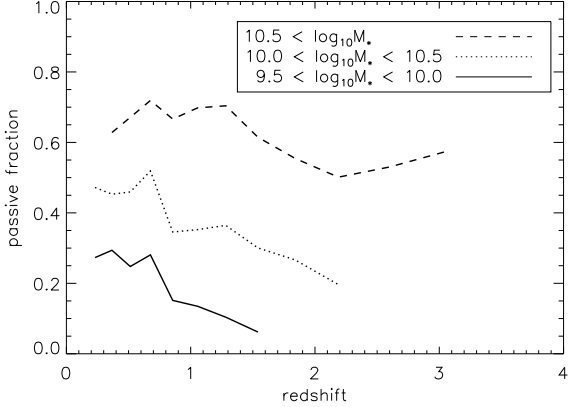}
\caption{The fraction of passive galaxies as a function of redshift for different stellar mass samples. Data are displayed only if they are above the $95\%$ completeness limit, and the weighted number of galaxies in the full sample exceeds 100. The peak visible at $z\sim0.65$ in the lower mass samples is due to a known cluster dominating the field at this redshift \protect\citep{vanBreuklen06}. Uncertainties are suppressed for clarity.}
\label{fracs}
\end{figure}

In Figure \ref{fracs} we plot the fraction of galaxies that are assigned to our passive galaxy sample as a function of redshift, separated by stellar mass. These fractions are derived from summing the galaxy weights described above in each redshift range for the passive samples and for the full sample. Samples in which the sum of galaxy weights in the combined passive plus star-forming sample is less than $100$ are not represented. Fractions are unreliable for small samples and we do not compute clustering measurements for these small samples. Furthermore only samples that are above the $95\%$ completeness limit are shown. There is a clear peak in the passive fraction of lower mass objects (log M$_{*}<10.5~$M$_{\odot}$) at the redshift corresponding to a massive cluster (or possible superstructure), at $z\sim0.65$ \citep{vanBreuklen06}. There is a possible second peak visible at $z\sim1.3$ which is as yet unconfirmed. The uncertainties on the fractions have been suppressed for clarity.

\section{Dark matter halo masses of passive and star-forming galaxies}
\label{method}

We now estimate the host dark matter halo masses of passive and star-forming sub-samples as a function of their stellar mass and redshift, beginning with high redshift ($2.00<z<3.36$ - an interval of $1500~$Mpc) passive galaxies. The typical mass of the parent halos that host a galaxy population can be estimated by measuring the linear bias of the population, with respect to the underlying dark matter density distribution. Such measurements can typically be achieved by computing the 2-point auto-correlation function (2pcf), $w(\theta)$, together with our knowledge of how the dark matter structure evolves with time (e.g. H10 and references therein). 

The 2pcf is the excess probability (over a random distribution) of finding a pair of galaxies at a given separation and can be estimated, using the \cite{Landy93} estimator, by
\begin{equation}
w(\theta)=\frac{N_{DD}(\theta)-2N_{DR}(\theta)+N_{RR}(\theta)}{N_{RR}(\theta)} + C.
\end{equation}
Here, $N(\theta)$ refers to the normalised number of pairs found at that angular separation and the subscripts DD, DR and RR represent pairs within the galaxy sample, pairs between the galaxy sample and a random catalogue and pairs within the random catalogue respectively. 

As described earlier, we use the probability of each galaxy lying within the desired redshift range as a weight. The pair counts are therefore the sum of the products of these weights over pairs of objects, rather than a simple pair count. The random catalogue must be sufficiently large such that statistical errors on $w(\theta)$ are dominated by real galaxy pairs, and must follow the same selection function as the galaxy sample. 
In the case of an angular correlation function for a sample of even depth, this simply means that the same mask must be used to remove bad regions in both the galaxy and random catalogues. 

Finally, the constant, $C$, represents a correction for the integral constraint, arising due to the limited field size. This constant depends on the intrinsic clustering strength and is computed, following \cite{Roche99}, using the random-random pairs as follows:
\begin{equation}
C=\frac{\Sigma N_{RR}(\theta).w(\theta)}{\Sigma N_{RR}(\theta)},
\end{equation}
As the integral constraint depends on the {\em intrinsic} clustering, rather than the observed clustering, it must be fit simultaneously with the bias factor (below). Uncertainties on $w(\theta)$ are estimated by a bootstrap analysis of $100$ repetitions.

In order to recover the host dark matter halo mass we must also know the angular correlation function of the dark matter distribution over the same volume as our galaxy sample. For this purpose we use the non-linear dark matter power spectrum routine of \cite{Smith03}, computed at the mean redshift of the sample. To obtain the non-linear dark matter angular correlation function, this is then Fourier transformed and projected using the full, weighted redshift probability distribution, p(z), of the galaxy sample (see section \ref{data} and Figure~\ref{pz}). Finally, we fit a single bias parameter to the galaxy clustering measurements,
\begin{equation}
w_{obs}(\theta) = b^2 \times w_{dm}(\theta),
\end{equation}
minimising the $\chi^2$. At large scales $w_{dm}$ is well described by linear gravity theory and the non-linear correlation function, $w_{\rm non-linear}$, is simply the same as the linear prediction, $w_{\rm linear}$. The assumption of linear bias is satisfied at these scales and a typical host halo mass for a galaxy sample can be estimated. 

At smaller scales non-linear effects become important, but by fitting the galaxy correlation function to the non-linear correlation function, using all points which satisfy $w_{\rm non-linear} < 2 \times w_{\rm linear}$, we can better constrain the fit. At scales smaller than this, non-linear effects dominate and we cannot assume that the galaxy distribution follows the dark matter distribution. The lower cut-off scale depends on the mean redshift of the galaxy sample, which for our $z>2$ samples described below corresponds to $15$ arcseconds. It can be seen in Figure \ref{wthplot} that the data follow the non-linear DM correlation function very well. This is unsurprising as \cite{Smith03} used the halo model \citep[see][for a review of the halo model]{Cooray02} to determine the dark matter power spectrum at scales where non-linear effects become important. This model has previously been used to describe galaxy clustering statistics with considerable success \citep[e.g.][]{Zheng07,Ross09}. We have verified that fits using non-linear scales in this way do not affect our conclusions, other than to reduce the uncertainties. 

At very large separations our measurements become unreliable, so we only use the data at separations up to $0.4$ degrees. In H10 we took an alternative approach, by fitting a power law with slope fixed at $\delta=-0.8$ to the galaxy angular correlation function which was then de-projected, from which we could infer a 3-dimensional correlation length. The advantage of projecting the dark matter correlation function, as we do here, is that we can avoid the possible degeneracies introduced by simultaneously fitting a power-law slope and the integral constraint.

\subsection{The cross-correlation function}
\label{crosscorrl}

A closely related measure to the 2pcf is the 2-point cross-correlation function (ccfn). The procedure is identical to that described for the 2pcf above, except that in the case of a ccfn we have two data sets, D1 and D2. The estimator is then modified as follows,
\begin{equation}
w(\theta)=\frac{N_{D1D2}(\theta)-N_{D1R}(\theta)-N_{D2R}(\theta)+N_{RR}(\theta)}{N_{RR}(\theta)}.
\end{equation}
Under the assumptions that both data sets trace the same dark matter distribution and are both linearly biased, we can use the bias measurement of a tracer population ($b_1$) with the ccfn bias ($b_{12}$) to infer the bias of the second population:
\begin{equation}
b_2 = \frac{b^2_{12}}{b_1}.
\end{equation}
Small sample sizes are one of the major difficulties that must be faced when using clustering to derive halo-mass estimates for passive galaxies at high redshift. Small samples lead to large statistical errors, dominating over other limiting factors such as sample variance, and consequently give very poor constraints on the halo mass. Due to the depth of our data we have a substantial population of galaxies at high redshift, though they are mostly star-forming objects. In using the cross-correlation statistics, we can cross a passive population for which we want to know the bias with a full mass-limited tracer population and achieve much smaller statistical uncertainties on their bias, $b_2$. 

Our weighted pair counts were made using a tree code adapted from Joshua Barnes' publicly available N-body gravity code \citep{Barnes86}. Using a tree rather than a brute force algorithm speeds up the computation markedly, but at the cost of precision. We ran a number of tests and determined the optimal opening angle of $\Theta=0.07$. Using this value the measurements are almost unaffected, however the bootstrap analysis is made significantly faster. 

\subsection{Clustering of passive galaxies at $z>2$}
\label{z2pass}

\begin{figure}
\includegraphics[angle=0, width=240pt]{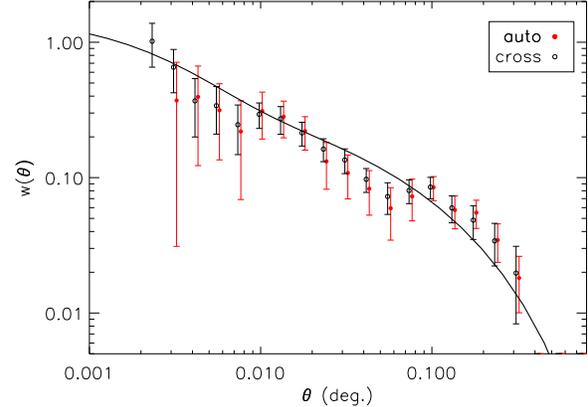}
\caption{Angular two point auto and cross-correlation functions for passive galaxies at $2.0<z<3.36$. The cross-correlation function is computed using the full $95\%$ mass-complete sample over the same redshift range. The cross-correlation function has been multiplied by ($b^2_{{\rm ccfn}}/b^2_{{\rm tracer}}$) to allow direct comparison with the auto-correlation function. The two sets of points agree very well with each other, but using a mass-limited tracer population to sample the dark matter distribution provides additional galaxy pairs which reduces the uncertainties in the cross-correlation function. Also shown is the dark matter angular correlation function, multiplied by the passive galaxy bias squared (solid line).}
\label{wthplot}
\end{figure}

\begin{figure}
\includegraphics[angle=0, width=240pt]{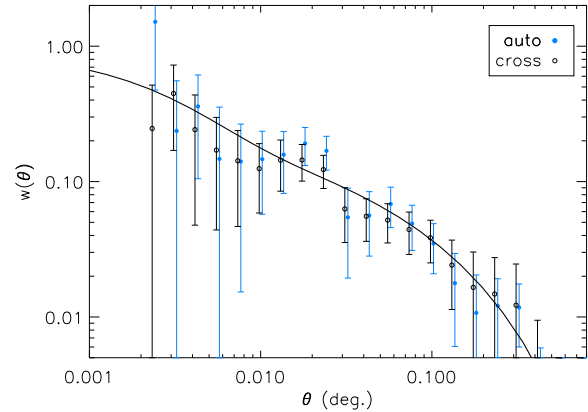}
\caption{Similar to Figure~\ref{wthplot} but for star-forming galaxies at the same redshift.}
\label{wthplotB}
\end{figure}

In H10 we suggested that if the observed clustering evolution of the passive galaxy population at $z<2$ were to continue at higher redshift, then passive and star-forming galaxies would be found in the same mass dark matter halos at $z\sim2.5$. Previously we had insufficient objects to obtain a meaningful measurement at this redshift. However, as the latest UDS release is more than 1 magnitude deeper than the data used in H10, we are now able to test this prediction. We use our passive galaxy sample, defined in Section \ref{data}, within the range $2.00<z<3.36$ (a depth of $1.5$ Gpc) and form a volume-limited sample by cutting at M$_*=10^{10.4}~$M$_{\odot}$. The sum of weights of passive objects with stellar mass, M$_* > 10^{10.4}~$M$_{\odot}$, within this redshift range, i.e. the expected number of galaxies, is $1202$. It is important to note that this number is dominated by objects with a high probability of lying in the redshift range $2.00<z<3.36$, i.e. with weight $>0.4$ ($\Sigma~ {\rm weight}_{0.4} = 1063$). In other words, the clustering measurements do not simply reflect a larger, low-redshift population with a small probability of being at $z>2$.

We compute measurements for both the 2pcf and ccfn, as detailed in the previous section. The tracer population used for the cross-correlation analysis was also constructed to be volume-limited, using all objects to the $95\%$ completeness limit at $z=3.36$ within the same redshift range as the target sample ($\Sigma~ {\rm weight} = 2450$, see section \ref{masses}). These measurements are shown in Figure \ref{wthplot} where red points are the 2pcf and black points are the ccfn. The bias of the tracer population has been multiplied out from the ccfn ($w^{\prime} = w \times b^2_{{\rm ccfn}}/b^2_{{\rm tracer}}$) to allow direct comparison with the 2pcf. Also shown is the non-linear dark matter angular correlation function multiplied by $b^2$, where $b$ is the bias of the passive sample implied by the ccfn analysis (black solid line). 

\begin{figure}
\includegraphics[angle=0, width=240pt]{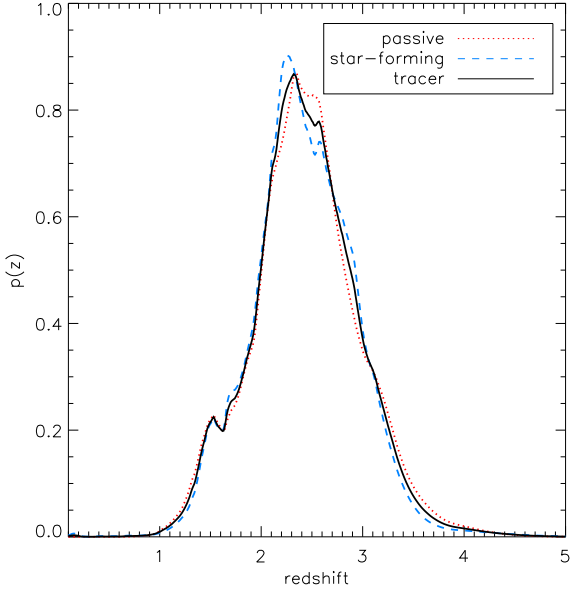}
\caption{Full weighted redshift probability distributions used to project the dark matter correlation function for our $z>2$ samples.}
\label{pz}
\end{figure}

As stated in section \ref{crosscorrl}, the cross-correlation technique relies on the target and tracer populations sampling the same dark matter structure. The redshift probability distributions that are used to project the dark matter correlation function are shown in Figure~\ref{pz}. The three distributions are very similar, giving us confidence that our analysis is robust.

Using the formalism of \cite{Mo02} we can directly link the galaxy bias to a typical host dark matter halo mass. The bias measured for this stellar-mass-limited passive galaxy sample suggests they are typically hosted by halos with mass, M$_{{\rm halo}}\sim 1.7\times10^{13}~$M$_{\odot}$ with bias, $b=5.23\pm 0.44$\footnote{ The bias values quoted here and elsewhere have been corrected for a systematic error which is detailed in section \ref{zbroad}.}, and an equivalent correlation length, $r_0 = 9.9\pm 0.9~$h$^{-1}$Mpc.

For comparison we plot the same measurements for star-forming galaxies in Figure~\ref{wthplotB}. The bias measured from this sample is $b=3.76\pm 0.35$, i.e. $\sim 2.5 \sigma$ lower than the passive galaxy sample. This bias implies halo masses of M$_{{\rm halo}}\sim 7.6\times10^{12}~$M$_{\odot}$ and correlation length, $r_0 = 7.3\pm 0.6~$h$^{-1}$Mpc.

For a stellar-mass-limited galaxy sample the host halo mass may also be recoved by using (sub-)halo matching, where the only important quatities are the space densities of the galaxy sample and their host halos \citep[e.g.][and references therein]{Moster10}. We can use this technique as a consistency check of our results. For the combined passive and star-forming samples the number density of galaxies is $2.5$ times the expected number density of halos with mass M$_{{\rm halo}}> 7.6\times10^{12}~$M$_{\odot}$, while for the passive sample the number density of galaxies is $\sim5$ times the expected number density of halos with M$_{{\rm halo}}> 1.7\times10^{13}~$M$_{\odot}$. Using UDS data, \cite{Quadri08} found an even greater disparity between the number densities of distant red galaxies (DRGs) and their host halos at similar redshifts. However, it was later shown by \cite{Tinker10a} that their results could be explained by a combination of satellite galaxies and sample variance. We discuss the effects of multiple occupation of halos in section \ref{HODeffects}.

\subsection{Evolution of host halo masses with redshift}
\label{zevol}

\begin{figure}
\includegraphics[angle=0, width=240pt]{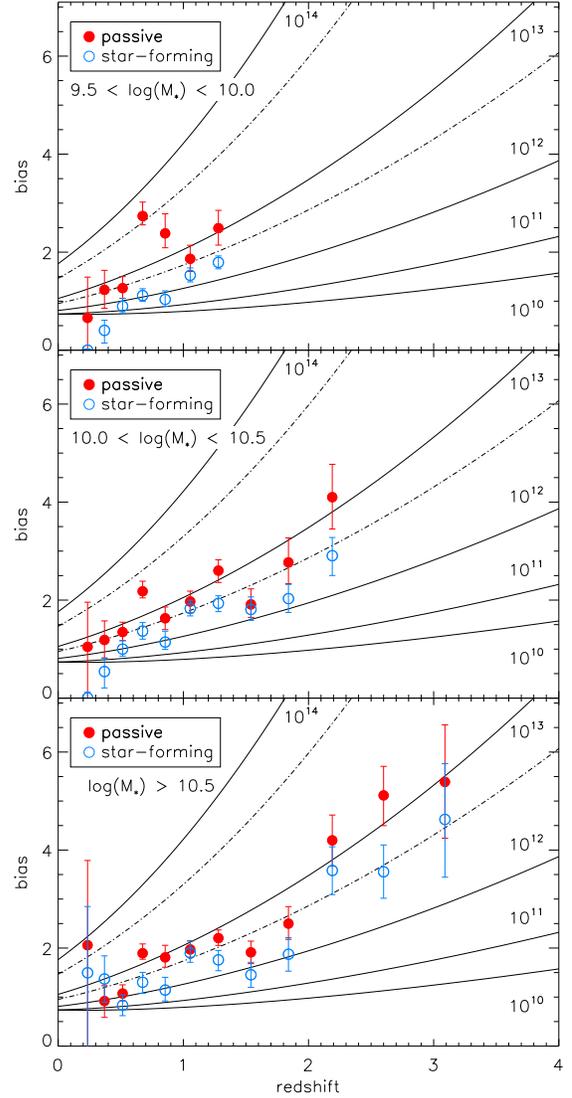}
\caption{Galaxy bias inferred from our cross-correlation analysis against redshift for passive and star-forming samples. In each panel the filled red points represent passive samples and the open blue points are star-forming samples. The stellar-mass range of galaxies in each panel is indicated in the upper left. Solid and dot-dashed lines show the bias for dark matter halos of various masses (in solar masses, as labelled). The dot-dashed lines are $5\times10^{13}~$M$_{\odot}$ and $5\times10^{12}~$M$_{\odot}$, upper and lower respectively.}
\label{UVJX}
\end{figure}

We now use the cross-correlation technique to perform a self-consistent analysis across all accessible redshifts ($0.3<z<3.36$). We split our sample in redshift with bins of width $500~$Mpc\footnote{This width is $\Delta~z\sim0.35$ at $z\sim2$ i.e. more than three times the expected photometric redshift dispersion} and in three mass intervals, $10^{9.5} < ~$M$/$M$_{\odot} < 10^{10.0}$, $10^{10.0} < ~$M$/$M$_{\odot} < 10^{10.5}$ and M$~>10^{10.5}$M$_{\odot}$ for both the passive and star-forming sample. As before, the tracer population is made up of all galaxies within the desired redshift range and with stellar mass greater than the $95\%$ completeness limit at the upper redshift bound. Any subsample in which the lower mass limit falls below the $95\%$ mass completeness limit is not considered. In Figure \ref{UVJX} we display the bias determined from our cross-correlation method as a function of redshift for passive and star-forming objects, with three panels representing the three bins in stellar mass. Red points correspond to our passive galaxy sample, while blue points represent the star-forming objects. Solid lines show how the clustering of dark matter halos depends on redshift at fixed halo mass (as labelled, in units of M$_{\odot}$) from the formalism given in \cite{Mo02} (and references therein). Finally, the dot-dashed lines are for two intermediate halo masses, $5\times10^{13}~$M$_{\odot}$ and $5\times10^{12}~$M$_{\odot}$ (upper and lower respectively).

From Figure \ref{UVJX} a number of trends are apparent. Firstly, passive galaxies are, on average, clearly more strongly clustered than their star-forming counterparts at similar stellar mass. This result is most apparent at lower stellar masses (M$~ <10^{10}~$M$_{\odot}$), which is driven in part by the structure at $z\sim0.65$ mentioned earlier. However, this behaviour remains even if we exclude the two affected redshift bins. The separation in clustering strength of passive and star-forming objects is less clear for the more massive galaxies (M$~ >10^{10}~$M$_{\odot}$), though still present. 
This difference is most pronounced for subsamples at $z<1$ which is consistent with H10, where we found a diminshing difference in clustering strength of passive and star-forming galaxies at higher redshift. Relative to H10, however, the extra information provided by the cross-correlation technique allows us to greatly reduce the width of the redshift intervals for our subsamples. Furthermore, we are able to obtain halo-mass estimates for high-mass passive objects at $z>2$ which was not previously possible.

A second feature of these results is the mass dependence within samples. The least massive of the passive galaxies appear to be the most strongly clustered (and therefore in the most massive dark matter halos). The bias typically decreases as the masses of the subsamples increase. This behaviour is clearest at $z<1$ and has previously been seen in \cite{McCracken08} and H10 at these redshifts. Despite this mass dependence, our passive galaxy samples appear to have consistently high parent halo masses, with typical masses M$_{\rm halo}\ge 5\times 10^{12}~$M$_{\odot}$ or higher. We note also the works of \cite{Chuter11} and \cite{Quadri12} who studied the small-scale environments of galaxy samples in the UDS field. They each found that low-mass or low-luminosity passive galaxies are found in the densest environments at any given redshift. 

For star-forming sub-samples, galaxies of different stellar masses but similar redshifts do not show any clear trends. The measurements are consistent with star-forming galaxies being typically hosted by similar mass halos irrespective of stellar mass. This absence of a clear correlation for star-forming galaxies was identified in the luminosity-selected subsamples in H10. Similarly, the downsizing-like behaviour we found in H10 is also present in this work, with sub-samples of equal stellar mass typically hosted by less massive halos at low redshift. Similar behaviour has previously been identified by \cite{Foucaud10} and \cite{Moster10} as well as H10. The bias evolution of passive galaxies with redshift at fixed stellar mass is not as clear as that seen in the star-forming sample, but nevertheless appears to be present. The implications of these findings are considered further in the following section.

\begin{figure*}
\includegraphics[angle=0, width=\textwidth]{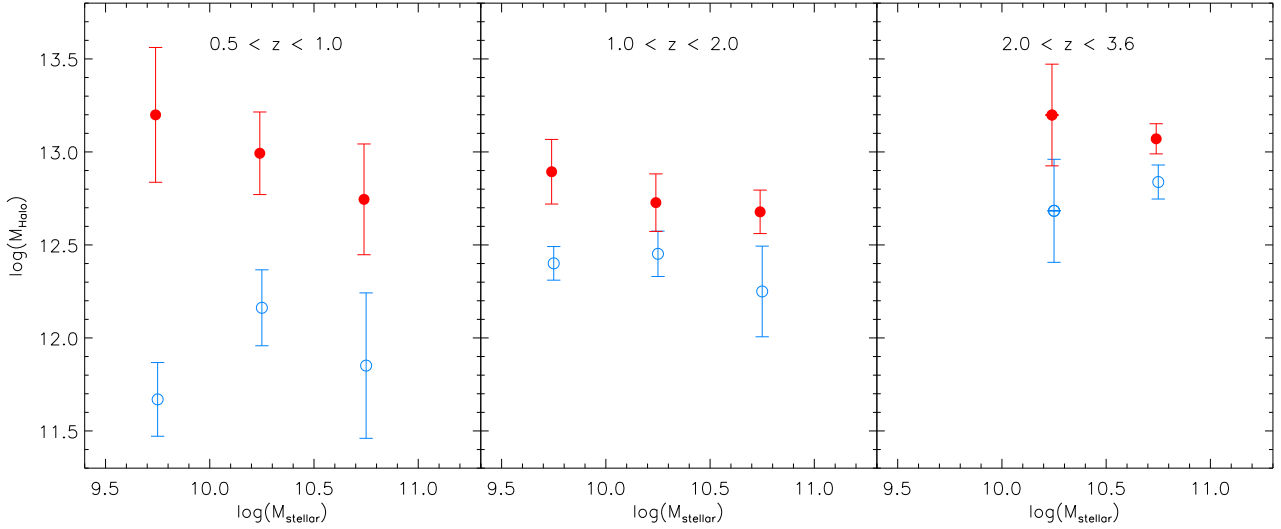}
\caption{Average halo masses implied by our bias measurements in three redshift intervals (as labelled). The data are mean halo masses for samples of fixed stellar mass, while the uncertainties are the standard error on the mean. As in previous Figures, filled red points represent passive galaxy samples and open blue points are for star-forming objects.}
\label{avgMz}
\end{figure*}

The raw measurements presented in Figure \ref{UVJX} are difficult to fully digest. We therefore condense our results into a more intuitive set of plots in Figure \ref{avgMz}. We plot the mean halo mass implied by our clustering results as a function of stellar mass for passive and star-forming subsamples in three redshift ranges: $0.5<z<1.0$, $1.0<z<2.0$ and $2.0<z<3.36$. Within each of these intervals we do not expect strong evolution in host halo mass (see \citealt{Conselice05} and \citealt{Brown08} for studies at $z<1$ and \citealt{Wake11} for a discussion of the $1<z<2$ interval) and so we can take a simple mean halo mass without biasing our interpretation. Each point in Figure \ref{UVJX} is converted to a typical halo mass, based on the measured bias, and these are then averaged over the indicated redshift intervals. The uncertainties on the data are the standard error on the mean. 

The behaviour described above can be seen more clearly in Figure \ref{avgMz}: the strong clustering of low-mass passive samples, the lack of a stellar-mass dependence on halo mass for star-forming galaxies and the `downsizing' of halo masses with redshift are all apparent. The averaged halo masses of all passive galaxy samples are greater than or equal to $5\times 10^{12}~$M$_{\odot}$ and within each redshift range the lowest-mass passive galaxies are found in the most massive halos. This result is clearest at low redshifts, indicating a build up of relatively low mass passive objects in high mass dark matter halos. Although there is no clear stellar-mass dependence on halo mass for star-forming galaxies, the downsizing-like behaviour is clearly apparent. Typical halo masses for star-forming galaxies in our lowest redshift interval are $\sim10^{12}$M$_{\odot}$, approaching $10^{13}~$M$_{\odot}$ at the highest redshifts.

\section[]{Discussion}
\label{disc}

\subsection{A halo-mass dependent process?}

We have tested whether a halo-mass dependent effect, such as the formation of a hot gaseous halo, is responsible for the establishment and build-up of the passive galaxy population. The hot halo favoured by theoretical models is expected to become efficient at terminating star formation when the halo mass exceeds M$_{\rm halo} > 10^{12.5}$M$_{\odot}$ by shock heating newly accreted gas and thereby preventing the gas supply available for star formation. Passive galaxies should therefore become much more abundant in halos of these masses. This in turn would result in a stronger clustering signal for passive galaxies over their star-forming counterparts which ought to be measureable in a data set such as the UDS. 

Taken at face value, the average parent halo masses for passive galaxy subsamples shown in Figure \ref{avgMz} are in agreement with a hot halo model and consistent with a limiting mass scale for passivity. Passive galaxies of all masses are typically hosted by halos of M$_{\rm halo} \ge 5\times 10^{12}~$M$_{\odot}$ or greater. The typical host halo masses for star-forming galaxies show no consistent stellar-mass dependence. However, at $z>2$ our star-forming samples are typically also found in halos with mass, M$_{\rm halo} \ge 5\times 10^{12}~$M$_{\odot}$. It is therefore possible that the efficiency with which a hot halo can shut-off star-formation evolves with redshift. 

A positive stellar-mass dependence on host halo mass is expected for galaxies at the centre of their dark matter halo, given that more massive dark matter halos will contain more gas available for star-formation (at least initially). The lack of a clear mass dependence within the star-forming sample is likely due to a combination of multiple occupation of halos (see below) and halo-mass dependent transition to the passive sample, supplemented by a possible contribution from sample variance. The inverse stellar-mass dependence shown by passive objects noted in the previous section could also result from halo occupation and a possible interpretation is discussed in detail below.

\subsection{Halo occupation effects}
\label{HODeffects}

Although our data support a halo-mass dependent process, they do not necessarily require it to be the hot halo that is responsible. We have so far neglected a number of alternative mass-dependent effects that may be occurring. In the idealised case where each dark matter halo hosts only a single object, the bias allows us to recover a minimum halo mass. However, this is only the case for the objects of highest stellar mass and so for the majority of our samples we recover an occupation-weighted bias, 
\begin{equation}
b=\langle{\rm N(M)}\rangle{\rm n(M)}b_{{\rm halo}}({\rm M})/n_g,
\end{equation}
where ${\rm n(M)}$ is the space density of halos of mass M, $\langle{\rm N(M)}\rangle$ is the mean number of galaxies residing in these halos, $b_{{\rm halo}}$ is the theoretical halo bias and $n_g$ is the space density of the galaxy sample \citep[e.g.][]{Cooray02}. Satellite galaxies are typically found in higher-mass halos than galaxies of the same stellar mass that lie at the centre of their halo. The higher the fraction of satellite galaxies in a sample, the further the measured typical host halo mass may be from the minimum mass of halos that host only central galaxies.

We may see the influence of halo occupation in the clustering results for the star-forming sample. Theoretical models predict a strong correlation between the stellar mass and host dark matter halo mass for the central galaxy of a halo. Our measurements do not discriminate between central and satellite galaxies and we find no clear stellar-mass dependence. These observed trends can be explained by an increased abundance of low stellar mass satellite galaxies in high mass halos that increase the measured bias. However, the action of a hot halo may also help explain the lack of a clear trend due to the significant scatter in the stellar to halo-mass relation of central galaxies \citep[e.g.][]{Yang09}. halo-mass dependent quenching would preferentially remove galaxies that inhabit more massive halos, leaving high stellar-mass galaxies that happen to be hosted by lower-mass halos in the star-forming sample. This process would naturally weaken any positive stellar to halo-mass relation for star-forming galaxies.
 
Given the possible influence of satellite galaxies we must consider the impact of satellite-related processes such as ram-pressure stripping \citep{Gunn72,Vollmer01}. Attempts have been made in the literature to do just this, using full halo occupation distribution modelling (HOD), where information on small scales is used to inform $\langle{\rm N(M)}\rangle$. Most notable for this discussion is the work by \cite{Tinker10b}. These authors used the angular clustering of \cite{Williams09} and found that they could construct an HOD which was marginally consistent with the data ($2\sigma$ from the best fit) and in which there was no difference in minimum halo mass between central passive and central star-forming galaxies at $1<z<2$. This is the behaviour that might be expected if a satellite quenching process was the dominant reason for the enhanced clustering of low-mass passive galaxies. As described below (section~\ref{Williams}), our samples differ from those of \cite{Williams09}, most notably in the redshift intervals considered. It remains to be seen whether the minimum halo masses of star-forming and passive central galaxies are similar for our samples. It is possible that there exist passive central galaxies in low-mass halos that are simply much less abundant than similar-mass satellite galaxies and some further investigation is required. Ideally we would perform a consistent HOD analysis of our data in narrow redshift ranges to minimise redshift effects, however we defer such an analysis to a later paper.

At low redshift it is possible to independently determine approximate halo masses and identify the central galaxy of a halo. Central galaxies should not be subject to stripping or other satellite processes and may provide a clearer picture. \cite{More11} used a group catalogue from the SDSS to compute average halo masses of passive and star-forming galaxies as a function of stellar mass, for central galaxies only. They find that for stellar masses, M$_{*}<10^{10.5}~$M$_{\odot}$, the average halo masses of passive galaxies are slightly greater than, but consistent with, those of the star-forming galaxies. However, at higher stellar masses differences become apparent. The dominant central population switches from star-forming to passive at M$_{{\rm halo}}\sim10^{12.3}~$M$_{\odot}$ and there are no star-forming central galaxies in halos with mass, M$_{{\rm halo}}>10^{12.8}~$M$_{\odot}$. This behaviour, and the mass scales at which it occurs, supports the halo-mass-scale process for ending star-formation; as does the absence of central passive galaxies in halos of M$_{{\rm halo}} < 10^{12}~$M$_{\odot}$ (though this latter point may be due also to incompleteness effects in the \citealt{More11} sample). Their transition mass scale (M$_{{\rm halo}}\sim10^{12.3}~$M$_{\odot}$) is lower than the scale we find for our lower redshift interval ($0.5<z<1$, M$_{{\rm halo}}\sim10^{12.7}~$M$_{\odot}$). This may further point towards a redshift dependence on halo-mass related quenching. However, given the differences in methodology, possible satellite effects and uncertainties, further work would be required to confirm any such evolution.

\cite{Peng10} also used SDSS data, claiming that they were able to separate the environmental quenching effects from effects due only to stellar mass. At fixed over-density they found that the environmental quenching rate did not depend on the stellar mass of galaxies considered. Ram pressure stripping might naively be expected to be more efficient in low-mass galaxies as they are less able to gravitationally bind their gas. This finding could then favour a strangulation-like process caused by the existence of a hot halo \citep{Larson80}.

Recently evidence has been found for a stronger correlation between a galaxy's star-forming or passive state and its structural parameters (e.g. Sersic index, velocity dispersion) than with its stellar mass \citep*{Wake12, Bell11}. This has been cited as evidence for a bulge-forming event being the key process in terminating star-formation, rather than a phenomenon due to stellar or halo mass. However, \cite{Wake12b} find that the host halo mass also correlates more closely with the central galaxy's structural parameters (particularly velocity dispersion) than stellar mass, so it is currently not clear that this argument must follow (though see \citealt{Li13} for a recent counter-argument). Furthermore, \cite{vanderWel11} and \cite{McLure12} find substantial fractions of passive disks at high redshift. Nevertheless, the development of a bulge is likely to be an important event in a galaxy's evolution. Perhaps bulge-forming events result in the expulsion of gas from a gas rich galaxy, but subsequent gas cooling and re-formation of a disk must prevented by a combination of environmental effects and possibly an AGN. 

\subsection[]{Comparison with previous work}

\subsubsection{Comparison with Williams et al. (2009)}
\label{Williams}

In \cite{Williams09} the authors found constant clustering strength with luminosity for their passive sample. In our present study we find a weak anti-correlation, with stronger clustering for lower-mass sub-samples. In order to investigate whether this apparent discrepancy can be explained by sample selection differences we have tested a simple UVJ selection as used in \cite{Williams09}. Using the same broad redshift interval ($1<z<2$) and an identical colour selection, but with our cross-correlation methodology, template fits and stellar masses, we recover the same behaviour. i.e. we find that the clustering strength does not depend on stellar mass within the passive sample. We suggest, therefore, that the build-up of passive satellites across this redshift range is appreciable and the use of narrow redshift intervals are important to identify subtle mass dependencies.

\subsubsection{Comparison with other literature results}

\begin{figure}
\includegraphics[angle=0, width=240pt]{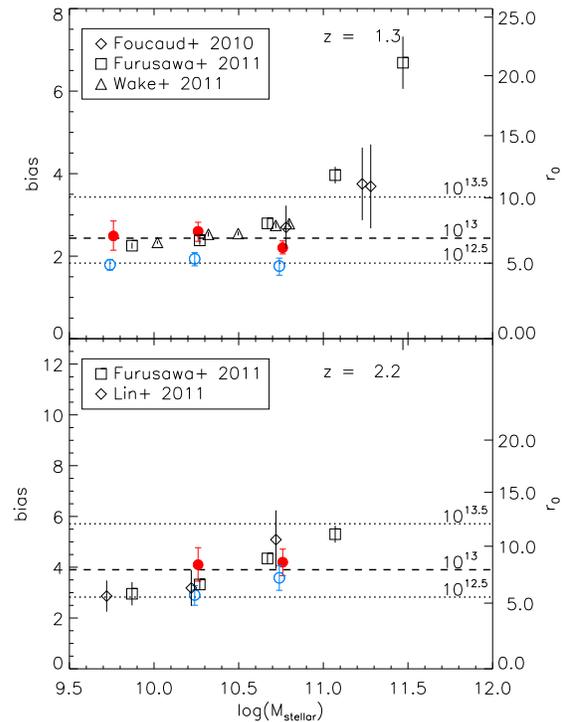}
\caption{Comparison of our results with measurements from the literature. We have chosen two of our redshift bins that coincide with a number of other studies. As in previous plots, red points correspond to passive subsamples and blue points are star-forming subsamples from our work. Horizontal lines show the effective bias of dark matter halos for three different masses (in solar masses) as labelled.}
\label{comparison}
\end{figure}

The literature on clustering measurements is extensive. The most directly comparable works are those of \cite{Meneux06, Coil08, McCracken08, Williams09, Foucaud10, Hartley10, Coupon11, Furusawa11, Wake11, Lin11} and \cite {Jullo12}, several of which we have mentioned previously. Many of these studies quote correlation lengths, $r_0$, while we report bias measurements. Following the assumptions made during computation of these $r_0$ values and our bias values, these are comparable quantities (though translating from one to the other is redshift dependent). Where appropriate we perform this conversion, using the \cite{Mo02} models used throughout this work.

In the following we shall consider two of our redshift intervals that coincide with the average redshifts of some recent results in the literature, at $z\sim1.3$ and $z\sim2.2$. Though we shall not cover results at $z<1$ in detail, our results are broadly consistent with the measurements of \cite{Meneux06, Coil08, McCracken08, Coupon11} and \cite{Jullo12}.

\subsubsection{Literature comparison at $z\sim1.3$}

In Figure~\ref{comparison} (upper panel) we plot our results at $z\sim1.3$ alongside those of \cite{Foucaud10, Furusawa11} and \cite{Wake11}. Each of these three studies consider only combined mass-selected samples, rather than the separate passive and star-forming samples that we do here. We therefore expect the literature measurements to lie between our points, but closer to the star-forming points at lower masses (as star-forming galaxies dominate by number at lower masses, see Figure~\ref{fracs}). Our results match the literature well except that for our highest-mass samples both passive and star-forming galaxies have slightly lower bias values than expected. Taking narrow redshift intervals helps to ensure that the underlying dark matter structure sampled by passive and star-forming objects is the same, but leaves our measurements slightly more susceptible to sample variance. It is most likely that sample variance is the cause of this discrepancy as it appears to affect both of our massive subsamples. Alternatively, it may be that there is a source of contamination that we have not been able to remove. Finally, we note that despite the fact that our redshift intervals are much narrower than those used for these literature results, the uncertainties on our bias measurements are similar or, in some cases, smaller due to the cross-correlation technique employed.

\subsubsection{Literature comparison at $z\sim2.2$}

The lower panel of Figure~\ref{comparison} shows a comparison of our $z\sim2.2$ results with two sets of measurements from the literature. As previously mentioned, \cite{Furusawa11} considered only mass-selected samples and these points should be compared with our star-forming subsamples. \cite{Lin11}, on the other hand, selected their objects by the $BzK$ technique \citep{Daddi04} and then split their star-forming $BzK$ sample by mass. Our measurements compare very well with these two literature results. We do not plot Lin et al.'s passive $BzK$ sample as the mean redshift is lower than their other points. However, they find passive objects to be very strongly clustered, more so even than our results, but with a substantial uncertainty.

\subsection[]{Discussion of potential biases}

The construction of our samples can be performed in a number of ways and although we believe we have used the optimal selection method for this study, there is no single correct choice. Whether we use a maximum likelihood redshift, minimum $\chi^2$ or the probabilistic approach taken here, the choice could potentially affect the results and interpretation. Similarly, our methodology could, in principle, impact upon our measurements. In this sub-section we briefly discuss the influence of our sample construction choices and implicit assumptions on our results.

\subsubsection[]{Redshift distributions}
\label{zbroad}

The redshift probability distribution functions, $p(z)$, that are produced by EAZY may not necessarily reflect the true probability distribution of any given galaxy. When these $p(z)$ are weighted and summed, the resulting redshift distribution of a sample that is used to project the dark matter may therefore differ from the intrinsic redshift distribution. This in turn may lead to systematic errors in the recovered bias values. With our spectroscopic sample we are able to test whether this is the case and apply a correction.

We split our spectroscopic sample into the same redshift intervals used in Figure \ref{avgMz}, using photometric redshifts. The $p(z)$ are summed in the same way as our science samples to form the photometric $n(z)$. This $n(z)$ and the true spectroscopic $n(z)$ for that subsample are each used to project an appropriate dark matter correlation function. The ratio of the two angular correlation functions that are produced in this way tells us whether we are over or underestimating the biases at these redshifts. We find that we require corrections to our final bias values by factors of $1.13$, $1.2$ and $1.4$ for $z<1$, $1<z<2$ and $z>2$ respectively. These are already included in all of our results and in the values given in the table at the end of this paper. In principle the required correction could depend not only on redshift, but also on SED and signal to noise (or equivalently, at a given redshift, stellar mass). However, our spectroscopic sample does not allow us to explore this possibility, and there is no existing sample which could be used to test such systematics for red galaxies at high redshift.

\subsubsection{Redshift correlations and contamination}
\label{zcorrlcont}

The second point to note is that, due to our use of a probabilistic redshift approach, the w$(\theta)$ measurements at different redshifts are correlated. This is true of any binning method as photometric redshift errors will scatter galaxies between redshift bins. However, there may be a concern that strongly clustered lower-redshift populations are driving high bias values at high redshift. We verify that the vast majority of our subsamples are dominated by objects that have a substantial probability ($p>0.4$) of being in the appropriate redshift range. The contamination by large numbers of galaxies that are at lower-redshifts should therefore be small. Excluding the small number of subsamples for which this is not strictly true from our analysis does not alter our conclusions. In addition to the above check, we repeat our measurements for samples that contain only galaxies with $p>0.4$, finding consistent behaviour. 

There is also a possibility that contamination by stars, cross-talk or bright AGN that have not been removed are affecting some measurements. We have investigated a number of different cuts to our catalogues: using the photometric redshift best-fit $\chi^2$, multi-wavelength data to remove possible AGN, different stellar selections and insisting upon a z-band detection for star-forming galaxies to remove potential cross-talk. Each set of measurements resulted in similar behaviour across the passive and star-forming samples. We present only those results for a fairly strict set of criteria (see section \ref{samplecons}) with the confidence that we have removed the majority of contaminants without biasing our results.

\subsubsection{IRAC PSF correction}

Our correction of the IRAC fluxes for the broad PSF and contamination by neighbouring sources makes the implicit assumption that both the source and the contaminant have the same K-3.6$\mu$m colour. If this assumption is false, then the IRAC fluxes could be inaccurate for a subset of our sources. In such cases, those objects may have poor fits during photometric redshift determination and erroneous stellar masses. The K-3.6$\mu$m colour of the full galaxy sample evolves with redshift, but is fairly constant at $z>1$. We test the impact of contaminants by comparing the redshift-dependent colour distribution of isolated galaxies with the broader population. These produce similar colour distributions, indicating that contamination is not a serious concern. Furthermore, as described above, we have reproduced our measurements using various cuts to remove contaminants, finding very similar results. We therefore expect that the number of objects which have close neighbours with very different K-3.6$\mu$m colours is small.

We further test the impact of the IRAC PSF correction on our redshifts by way of a monte-carlo analysis. For two of our samples, the $2.0<z<3.36$ passive sample used in section \ref{z2pass} and the $1.68<z<2.0$ star-forming sample, we randomly vary the correction to the IRAC fluxes by between zero and twice the computed correction. The passive sample at $z>2$ is important as it is the first such sample for which a clustering measurement has been made. The $1.68<z<2.0$ star-forming sample is likely to be among the most difficult to achieve an accurate redshift as the $4000{\rm \AA}$ break is in the Y-band where we have no data and the Lyman limit is still blue-ward of our u-band. The typical uncertainties in the weights that are used in our cross-correlation analysis are $\sim 4.7\%$ and $\sim 9.4\%$ for these two samples respectively. At these levels we do not expect any inaccuracies in the PSF correction to impact upon our conclusions. 

\subsubsection{Stellar mass uncertainties}

During stellar mass and star formation rate estimation we only considered templates with exponentially declining star formation rates. This approach is common in the literature, but it is becoming increasingly apparent that there is a population of objects at high-redshift which are better fit by an exponentially increasing star-formation history \citep[e.g.][]{Papovich11}. Our main concerns are the separation of our sample into passive and star-forming objects and their stellar masses. Clearly, not considering increasing star-formation rates will have no impact upon their sample classification. However, the derived stellar masses for some high-redshift star-forming galaxies may be affected by our assumption, resulting in over-estimated stellar masses for these galaxies. Our most massive passive and star-forming high-redshift subsamples already have comparable clustering strength. If a number of the star-forming objects are of intrinsically lower mass, then the clustering of the true high-mass star-forming subsample should be even higher. It would be unexpected, but not impossible, that high-mass star-forming galaxies are in more massive halos than their passive counterparts. We anticipate, however, that any effect such as this is minor. 

Our stellar masses have uncertainties associated with them arising from photometric precision, template fitting and redshift tolerance within a bin. These uncertainties are mitigated somewhat by our probabilistic weighting approach, but we cannot rule out cross-contamination between our stellar-mass bins. The uncertainty due to template fitting alone has been estimated by \cite{Bolzonella10} for 12 bands of photometry covering the u-band to near-infrared at better than $\sim0.2$ dex, based on simulated catalogues. This uncertainty reflects the variation in mass-to-light ratio that a galaxy may have. In addition, the luminosity of a galaxy is uncertain. Based on the breadth of our redshift bins, we can compute a maximum uncertainty on the luminosity of an object. Converting this to an error on the mass we find an average $\sim0.2$ dex with slightly greater error, up to $0.3$ dex, at low redshift and slightly lower, $0.1$ dex, at our highest redshifts. Added in quadrature we find a final typical uncertainty of $\sim0.3$ dex on our stellar masses, significantly smaller than our $0.5$ dex bin width but nevertheless a possible cause of cross contamination. Such contamination will tend to reduce any clustering dependence on stellar mass and therefore strengthens our conclusion regarding the stellar-mass dependent clustering of passive galaxies. 

\subsubsection{Sample variance}

In our error estimate we do not take sample variance into account as this is best accomplished by comparing our results with independent fields. Our binning in redshift corresponds to only $500~$Mpc in depth and the influence of structure is clearly visible in Figure \ref{UVJX}. Within a redshift interval, however, both star-forming and passive objects are subject to the same structure and so their bias measurements should be similarly affected. Although our uncertainties are under estimated, the comparison of passive and star-forming objects should therefore be reasonable.

\section[]{Conclusions}
\label{conc}

We have used a cross-correlation function analysis to estimate the host halo masses of star-forming and passive galaxies extracted from the UKIDSS Ultra Deep Survey to $z=3.36$. Relative to an auto-correlation function, our cross-correlation analysis provides much tighter constraints on the typical host halo masses of rare galaxy samples. This is due to the use of a much more abundant tracer population to sample the dark matter distribution. Passive galaxies were identified initially from a rest-frame colour selection which was then refined using template fits and $24\mu$m data. The star-forming sample was made up of the remainder of the parent sample. Our main findings are as follows: 

We presented the first estimate of the typical host dark matter halo masses for a stellar-mass-selected (M$_{{\rm *}} > 10^{10.4}~$M$_{\odot}$) passive galaxy sample at $z>2$. This epoch is important as it is here that the first substantial populations of passive galaxies are found. Our measurements imply a typical host halo mass of M$_{{\rm halo}}>1.5\times10^{13}~$M$_{\odot}$, which is higher than those of equal stellar mass $z>2$ star-forming galaxies at a $\sim2.5\sigma$ significance level.

We also performed a detailed analysis of our star-forming and passive galaxies in differential subsamples, $500$Mpc in depth and in three stellar-mass ranges: $10^{9.5} < {\rm M}_{{\rm *}}/{\rm M}_{\odot}< 10^{10.0}$, $10^{10.0} < {\rm M}_{{\rm *}}/{\rm M}_{\odot}< 10^{10.5}$ and M$_{{\rm *}}/{\rm M}_{\odot} > 10^{10.5}$. We confirmed many of the previous findings from \cite{Hartley10} but over a wider range in redshift and for much narrower redshift intervals. The notable results from this analysis are:
\begin{itemize}
\item The mass of the host dark matter halos are clearly higher for passive galaxies than star-forming objects.

\item  The difference in host halo masses of passive and star-forming galaxy samples diminishes at higher stellar mass (M$_{*}>10^{10.0}~$M$_{\odot}$) and at higher redshift.

\item All passive galaxies are hosted by high-mass halos. The measurements are consistent with a mechanism for galaxy transition to passivity requiring host halo masses, M $\sim 5\times 10^{12}~$M$_{\odot}$. However, the typical host halo masses of star-forming galaxies at $z>2$ are also above this limit. This may indicate a more complex situation in which the efficiency of quenching by a hot halo is redshift dependent.

\item Low-mass passive galaxies (M$_{*}<10^{10.0}~$M$_{\odot}$) are hosted by the most massive dark matter halos. This counter-intuitive behaviour is most likely due to efficient processing of low-mass satellite galaxies in high mass halos and implies that multiple mechanisms may be involved in terminating star-formation.

\item The host halo masses of star-forming galaxies, on the other hand, show little dependence on stellar mass. We interpret this relative absence of a correlation as being due to a delicate balance between two factors: the presence of a mild positive correlation between stellar and halo masses for central galaxies and the greater abundance of low-mass satellite galaxies in more massive dark matter halos. 
\end{itemize}

The results we have presented are an important step in determining which process (or processes) are responsible for the termination of star-formation in galaxies. Our results indicate that a process dependent on halo mass is important for galaxies of all masses, but that it is likely supplemented by a process that is efficient in processing low-mass satellite systems, such as ram-pressure stripping. 

\section*{Acknowledgments}
We wish to extend our deepest gratitude to the staff at UKIRT for their tireless efforts in ensuring the sucess of UKIDSS. WGH would also like to acknowledge the support of the STFC during the preparation of this work. RJM acknowledges Royal Society funding through the award of a University Research Fellowship. JSD acknowledges the support of the Royal Society through a Wolfson Research Merit award. Finally, we'd like to thank the annonymous referee for their detailed reading of our manuscript.

%\section*{Appendix}
\begin{table*}
\centering
\begin{tabular}{|l|l|l|l|l|l|l|l|l|l|l|}
N$_{{\rm obj}}$&$\Sigma$~weight&z$_{{\rm min}}$&z$_{{\rm max}}$&$\bar{{\rm z}}$&M$_{{\rm min}}$&M$_{{\rm max}}$&b&$\sigma_{{\rm b}}$&$r_{0}^{*}$&$\sigma_{r_0}$\\
\hline
{\bf passive}& & & & & & & & & & \\
    165 &   63.05 &    0.17 &    0.30 &    0.26 &   10.00 &   10.50 &    1.05 &    1.08 &    5.34 &    5.01\\
     53 &   22.57 &    0.17 &    0.30 &    0.26 &   10.50 &   12.60 &    2.06 &    1.92 &   11.01 &   12.10\\
    139 &   44.41 &    0.17 &    0.30 &    0.26 &    9.50 &   10.00 &    0.66 &    1.13 &    0.00 &    3.98\\
    387 &  183.31 &    0.30 &    0.44 &    0.39 &   10.00 &   10.50 &    1.19 &    0.38 &    5.44 &    2.01\\
    260 &  133.17 &    0.30 &    0.44 &    0.41 &   10.50 &   12.60 &    0.92 &    0.32 &    4.04 &    2.86\\
    289 &  148.42 &    0.30 &    0.44 &    0.38 &    9.50 &   10.00 &    1.23 &    0.39 &    5.74 &    1.75\\
    856 &  405.30 &    0.44 &    0.59 &    0.54 &   10.00 &   10.50 &    1.35 &    0.19 &    5.70 &    1.12\\
    898 &  483.30 &    0.44 &    0.59 &    0.53 &   10.50 &   12.60 &    1.07 &    0.16 &    4.30 &    0.71\\
    556 &  246.51 &    0.44 &    0.59 &    0.54 &    9.50 &   10.00 &    1.27 &    0.22 &    5.02 &    0.96\\
    912 &  400.83 &    0.59 &    0.76 &    0.67 &   10.00 &   10.50 &    2.18 &    0.17 &    8.61 &    0.50\\
   1105 &  519.38 &    0.59 &    0.76 &    0.68 &   10.50 &   12.60 &    1.89 &    0.16 &    6.96 &    0.80\\
    630 &  254.08 &    0.59 &    0.76 &    0.68 &    9.50 &   10.00 &    2.74 &    0.23 &   11.18 &    0.80\\
    820 &  368.58 &    0.76 &    0.95 &    0.87 &   10.00 &   10.50 &    1.63 &    0.23 &    5.66 &    0.67\\
   1254 &  640.44 &    0.76 &    0.95 &    0.87 &   10.50 &   12.60 &    1.81 &    0.22 &    6.39 &    0.73\\
    556 &  229.68 &    0.76 &    0.95 &    0.87 &    9.50 &   10.00 &    2.38 &    0.35 &    8.08 &    1.88\\
   1194 &  445.23 &    0.95 &    1.16 &    1.07 &   10.00 &   10.50 &    1.97 &    0.22 &    5.94 &    0.76\\
   1808 &  805.42 &    0.95 &    1.16 &    1.06 &   10.50 &   12.60 &    1.97 &    0.17 &    5.94 &    0.62\\
    659 &  210.05 &    0.95 &    1.16 &    1.07 &    9.50 &   10.00 &    1.86 &    0.27 &    5.59 &    1.18\\
   1563 &  490.98 &    1.16 &    1.40 &    1.33 &   10.00 &   10.50 &    2.60 &    0.23 &    7.13 &    0.31\\
   2003 &  876.19 &    1.16 &    1.40 &    1.30 &   10.50 &   12.60 &    2.20 &    0.16 &    6.28 &    0.33\\
    756 &  171.36 &    1.16 &    1.40 &    1.32 &    9.50 &   10.00 &    2.49 &    0.35 &    7.21 &    1.31\\
   1254 &  419.91 &    1.40 &    1.68 &    1.52 &   10.00 &   10.50 &    1.91 &    0.29 &    4.54 &    0.81\\
   1939 &  840.84 &    1.40 &    1.68 &    1.53 &   10.50 &   12.60 &    1.91 &    0.23 &    4.68 &    0.67\\
   1137 &  273.86 &    1.68 &    2.00 &    1.87 &   10.00 &   10.50 &    2.77 &    0.47 &    6.19 &    1.44\\
   1786 &  555.68 &    1.68 &    2.00 &    1.82 &   10.50 &   12.60 &    2.50 &    0.33 &    5.87 &    1.00\\
    971 &  204.03 &    2.00 &    2.38 &    2.19 &   10.00 &   10.50 &    4.10 &    0.66 &    9.18 &    1.94\\
   1594 &  404.59 &    2.00 &    2.38 &    2.17 &   10.50 &   12.60 &    4.20 &    0.52 &    8.66 &    1.65\\
    929 &  351.86 &    2.38 &    2.82 &    2.63 &   10.50 &   12.60 &    5.11 &    0.60 &    9.64 &    1.05\\
    948 &  230.68 &    2.82 &    3.36 &    3.03 &   10.50 &   12.60 &    5.39 &    1.16 &    8.29 &    2.17\\
\hline
{\bf star forming}& & & & & & & & & & \\
    172 &   70.61 &    0.17 &    0.30 &    0.26 &   10.00 &   10.50 &    0.00 &    0.06 &    0.00 &    0.00\\
     66 &   25.75 &    0.17 &    0.30 &    0.26 &   10.50 &   12.60 &    1.50 &    1.54 &    7.93 &    7.48\\
    345 &  118.17 &    0.17 &    0.30 &    0.25 &    9.50 &   10.00 &    0.00 &    0.01 &    0.00 &    0.00\\
    546 &  221.45 &    0.30 &    0.44 &    0.39 &   10.00 &   10.50 &    0.54 &    0.30 &    0.00 &    1.76\\
    175 &   78.87 &    0.30 &    0.44 &    0.38 &   10.50 &   12.60 &    1.37 &    0.47 &    6.25 &    2.13\\
    836 &  356.76 &    0.30 &    0.44 &    0.39 &    9.50 &   10.00 &    0.40 &    0.23 &    0.00 &    0.00\\
    818 &  477.01 &    0.44 &    0.59 &    0.52 &   10.00 &   10.50 &    1.00 &    0.16 &    4.25 &    0.72\\
    399 &  236.92 &    0.44 &    0.59 &    0.51 &   10.50 &   12.60 &    0.83 &    0.21 &    3.30 &    2.17\\
   1484 &  748.25 &    0.44 &    0.59 &    0.53 &    9.50 &   10.00 &    0.90 &    0.15 &    3.42 &    0.58\\
   1057 &  371.74 &    0.59 &    0.76 &    0.69 &   10.00 &   10.50 &    1.37 &    0.17 &    5.02 &    0.84\\
    517 &  202.67 &    0.59 &    0.76 &    0.69 &   10.50 &   12.60 &    1.30 &    0.21 &    4.81 &    0.84\\
   1898 &  650.91 &    0.59 &    0.76 &    0.69 &    9.50 &   10.00 &    1.11 &    0.13 &    4.05 &    0.56\\
   1629 &  695.72 &    0.76 &    0.95 &    0.89 &   10.00 &   10.50 &    1.14 &    0.19 &    3.65 &    0.70\\
    674 &  323.45 &    0.76 &    0.95 &    0.87 &   10.50 &   12.60 &    1.14 &    0.24 &    3.86 &    0.94\\
   2993 & 1241.7 &    0.76 &    0.95 &    0.89 &    9.50 &   10.00 &    1.04 &    0.16 &    3.29 &    0.62\\
   2227 &  821.83 &    0.95 &    1.16 &    1.06 &   10.00 &   10.50 &    1.83 &    0.16 &    5.71 &    0.84\\
    820 &  349.90 &    0.95 &    1.16 &    1.05 &   10.50 &   12.60 &    1.90 &    0.22 &    6.19 &    0.78\\
   4086 & 1372.3 &    0.95 &    1.16 &    1.07 &    9.50 &   10.00 &    1.53 &    0.14 &    4.46 &    0.46\\
   2743 &  856.55 &    1.16 &    1.40 &    1.33 &   10.00 &   10.50 &    1.93 &    0.16 &    5.16 &    0.40\\
   1018 &  368.56 &    1.16 &    1.40 &    1.33 &   10.50 &   12.60 &    1.76 &    0.21 &    4.55 &    0.62\\
   5241 & 1479.5 &    1.16 &    1.40 &    1.33 &    9.50 &   10.00 &    1.79 &    0.13 &    5.00 &    0.42\\
   2716 &  979.46 &    1.40 &    1.68 &    1.53 &   10.00 &   10.50 &    1.81 &    0.24 &    4.44 &    0.54\\
   1340 &  526.74 &    1.40 &    1.68 &    1.54 &   10.50 &   12.60 &    1.45 &    0.25 &    3.41 &    0.72\\
   2624 &  724.40 &    1.68 &    2.00 &    1.82 &   10.00 &   10.50 &    2.03 &    0.29 &    4.51 &    0.55\\
   1522 &  423.67 &    1.68 &    2.00 &    1.83 &   10.50 &   12.60 &    1.87 &    0.34 &    4.05 &    0.68\\
   3425 &  815.65 &    2.00 &    2.38 &    2.16 &   10.00 &   10.50 &    2.91 &    0.39 &    6.17 &    1.13\\
   1669 &  416.27 &    2.00 &    2.38 &    2.16 &   10.50 &   12.60 &    3.58 &    0.49 &    7.55 &    1.17\\
    995 &  311.84 &    2.38 &    2.82 &    2.58 &   10.50 &   12.60 &    3.56 &    0.54 &    6.44 &    1.13\\
    679 &  173.68 &    2.82 &    3.36 &    2.98 &   10.50 &   12.60 &    4.63 &    1.16 &    8.14 &    2.49\\
\hline
\end{tabular}
\caption{Table of values used in Figure \ref{UVJX}. Columns are: number of objects, sum of weights (i.e. the expected number of galaxies, see section \ref{samplecons}), redshift interval and median redshift, stellar-mass interval, bias, $1\sigma$ uncertainty on the bias and the corresponding $r_0$ value and uncertainty. 
$^*$Note that $r_0$ values are computed from the halo models of \protect\cite{Mo02}, and where the bias corresponds to an unphysical halo mass, a value of zero is used.}
\end{table*}

\bibliographystyle{mn2e.bst}
\bibliography{mn-jour,papers_cited_by_WH}

\label{lastpage}

\end{document}